# Dynamics of Vortex-Induced-Vibrations of a Slit-Offset Circular Cylinder for Energy Harvesting at Low Reynolds Number

Mayank Verma[1], Ashoke De[1,2 a)]

[1]*Department of Aerospace Engineering, Indian Institute of Technology Kanpur, Kanpur, 208016, India.*

[2]*Department of Sustainable Energy Engineering, Indian Institute of Technology Kanpur, Kanpur, 208016, India.*

Vortex-Induced Vibrations (VIV) offer a safe, renewable, and environmentally friendly energy source for energy harvesting. To enhance the energy harvesting capability of the circular cylinder-based devices, the authors explore the placement of the normal slit by determining the most effective slit offset location from the cylinder's center. Using the open-source Computational Fluid Dynamics (CFD) program OpenFOAM, a series of numerical simulations are conducted to determine the utility of the slit placement on the circular cylinder and how it influences (positively/negatively) the harvester's performance for a 1-degree-of-freedom (1-DOF) VIV system. The study shows that the slit-cylinder displays the three-branch VIV response (i.e. initial branch (IB), upper branch (UB), and lower branch (LB)). At a Reynolds number of 150, the cylinder with no slit exhibits the two-branch VIV response (i.e. IB and LB) but lacks the upper branch. This is the first time the Upper Branch has been acquired for such low Re flows. The results indicate that adding a normal slit to the middle of the cylinder improves the alternating suction and blowing phenomena. Offsetting it from the center towards the back stagnation point suppresses the VIV and makes it unsuitable for energy harvesting applications. While positioning the slit toward the front stagnation point improves aerodynamic lift, and the cylinder sheds vortex closer to each other. It enhances the amplitude of transverse oscillation and, consequently, the power extraction. In addition, the peak energy transfer ratio for these scenarios is comparable to that of the no-slit case but with a larger range of peak energy transfer ratio values. It makes it suitable for energy harvesting applications.

*Nomenclature*

| | | |
|---|---|---|
| L | = | Length, m |
| D | = | Diameter of the cylinder, m |
| s | = | Slit width, m |
| Re | = | Reynolds number, $UD/\upsilon$ |
| m | = | Total mass of the system (mass of the cylinder and added mass), kg |
| $m^*$ | = | Mass ratio, $\dfrac{4m}{\pi\rho D^2 L}$ |
| p | = | Static pressure, Pa |
| k | = | Spring constant of the system, N/m |
| $k^*$ | = | non-dimensional spring constant, $\dfrac{k}{\rho U_\infty^2 L}$ |
| c | = | Damping of the system, N-s/m |
| $C^*$ | = | non-dimensional damping coefficient, $\dfrac{c}{\rho U_\infty DL}$ |
| $f_n$ | = | Natural frequency of the system, Hz |
| $f$ | = | Non-dimensional oscillation frequency, $\dfrac{f_s D}{U}$ |
| $f^*$ | = | Normalized oscillation frequency, $f/f_n$ |
| $U_r$ | = | Non-dimensional reduced velocity, $\dfrac{U_\infty}{f_n D}$ |

___________________________

a) Author to whom correspondence should be addressed. Electronic mail: ashoke@iitk.ac.in

| | | |
|---|---|---|
| $A_y^*$ | = | Non-dimensional transverse oscillation amplitude, $\dfrac{(Y/D)_{max} + abs(Y/D)_{min}}{2}$ |
| z | = | Cell displacement field, m |
| l | = | Cell center distance, m |
| P(t) | = | Instantaneous extracted power, $\dfrac{2c\dot{y}^{*2}}{\rho U^3 DL}$ |
| $P_{avg}$ | = | Average extracted power, $4\pi^4 m^* \zeta \dfrac{((f_s/f_n)A_Y^*)^2}{U_r^3}$ |
| $P_{fluid}$ | = | Available fluid power, $\dfrac{1}{2}\rho U^3 DL(1+2A_Y^*)$ |
| $C_L$ | = | Aerodynamic Lift coefficient |
| $C_D$ | = | Aerodynamic Drag Coefficient |
| h | = | Grid spacing |
| r | = | Grid refinement ratio, $r_{i+1,i} = \dfrac{h_{i+1}}{h_i}$ |
| E | = | Error based on the Richardson Error estimator |
| o | = | Order of convergence |
| N | = | Total number of grid points |

*Greek Symbols*

| | | |
|---|---|---|
| $\alpha$ | = | Slit offset angle, degree |
| $\rho$ | = | Fluid density, kg/m³ |
| $\mu$ | = | Dynamic viscosity of the fluid, kg-m/s |
| $\vec{v}$ | = | Velocity vector, m/s |
| $\upsilon$ | = | Kinematic viscosity, m²/s |
| $\zeta$ | = | Damping coefficient, $\zeta = \dfrac{c}{2\sqrt{km}}$ |
| $\omega_n$ | = | Natural frequency of the system, $\omega_n = \sqrt{\dfrac{k}{m}}$ |
| $\tau$ | = | Non-dimensional time, $\tau = \dfrac{tU}{D}$ |
| $\gamma_m$ | = | Mesh diffusion coefficient |
| $\eta$ | = | Energy transfer ratio, $\dfrac{P_{avg}}{(1+2A_Y^*)}$ |
| $\varepsilon$ | = | Error, $\varepsilon_{i+1,i} = \dfrac{f_{i+1} - f_i}{f_i}$ |
| $\beta$ | = | Slit inclination angle, degree |

*Subscripts*

| | | |
|---|---|---|
| u | = | Upstream direction |
| d | = | Downstream direction |
| h | = | Height |
| n | = | Natural value |
| ∞ | = | Freestream value |
| s | = | Shedding |
| Y | = | Transverse direction |
| max | = | Maximum value |
| min | = | Minimum value |
| m | = | Mesh |
| rms | = | r.m.s. value |
| mean | = | Mean value |

# I. INTRODUCTION

Flow over a circular cylinder is a classic case of fluid mechanics widely studied due to its numerous engineering applications and rich fluid physics. When fluid flows over the cylinder, alternate vortex shedding begins (if the Reynolds number exceeds a particular value). This periodic vortex loss modifies the base pressure and causes the varying lift force on the cylinder. If the body is elastically mounted (or free to oscillate), it will exhibit structural vibrations and generate acoustic noise[1, 2]. When the frequency of these structural vibrations equals the natural frequency of the system, a phenomenon known as "lock-in" in the scientific literature[3-5], the cylinder oscillates with large oscillation amplitudes. These high-amplitude oscillations may cause the structure to break catastrophically. Consequently, a significant proportion of VIV research is devoted to discovering effective techniques for mitigating vibrations, such as suction and blowing[6], momentum injection[7], splitter plates[8], etc.

In contrast, these high-amplitude oscillations can be employed effectively for various applications, including mixing augmentation, energy harvesting, vortex flowmeters, etc. A slit in the cylinder provides a passive mechanism for controlling the flow based on its orientation. This article studies the fluid dynamic properties of an elastically mounted circular cylinder with a normal slit and its effect on oscillation amplitude and fluid-flow distribution when the slit is offset from the center of the cylinder.

The use of slits to modify the flow behind the cylinder dates back to 1978 when Igarashi[9] investigated the vortex shedding behind a cylinder with a slit at a Reynolds number of $\mathrm{Re} = 1.38 - 5.20 \times 10^4$. This remarkable study described two flow control methods using a slit: wake control by self-injection (for the slit inclination angle $0^o \leq \beta \leq 40^o$) and the alternate boundary layer suction and blowing (for $60^o \leq \beta \leq 90^o$). They also reported the potential and separated flow behind the slit caused by boundary layer suction and blowing, respectively. They concluded that the wake of the cylinder with a normal slit ($\beta = 90^o$) resembles that of the oscillating aerofoil[10]. This prompted the authors to select the inclination angle of $90^o$ for this study.

Further, Popiel et al.[11] introduced the normal slit with a concave rear surface and observed more stable and stronger vortex shedding, making it the best vortex shedder available. Olsen & Rajagopalan[12] also reported the increased strength of the vortices for the normal slit case. Recently, many studies[13-18] have looked at the effect of the slit placed parallel to the incoming flow and demonstrated the efficient VIV suppression. Ma and Kuo[19] showed that the pressure difference between the slit openings is the driving force for the periodic suction/blowing with zero-net-mass flux. Recently, Belamadi et al.[20] used slits in the aerofoils of the wind turbine blades to improve the aerodynamic performance. Similarly, Kim et al.[21] and Watanable & Ohya[22] studied the VIV phenomena for wind and water turbines. In the experimental study of a normal slit cylinder $Re \approx 2\times10^3 - 5\times10^4$, Peng et al.[23] reported the linear relation of the Strouhal number against the Reynolds number. Recently, Zhu et al.[24] conducted an experimental study to understand the effect of offsetting the normal slit (keeping the cylinder stationary) from the cylinder's center $Re \approx 2 - 5.4\times10^3$. Their study demonstrates that the slit-vent produces a strong effect of blowing and suction on the boundary layer if placed near the rear-stagnation point of the cylinder.

Notably, the slit-cylinder is studied majorly for the high Reynolds numbers. In reality, the flow over the structures may experience a wide range of Re, especially in the water flow where the viscosity is very high. Also, a thorough understanding of the flow distribution, including the boundary layer development, is vital for modifying the cylinder with the slit. However, most of the studies available deal with the fixed/stationary slit-cylinder. Very few studies (Baek & Karniadakis[25], Verma et al.[18]) considered the actual motion of the slit-cylinder. The intent of those studied was to suppress the VIV. Deriving the motivation from this, the authors performed this study to characterize the fluid flow better to develop a thorough understanding of the flow physics associated with the oscillations of the slit-cylinder under 1-DOF VIV at a low Reynolds number. Therefore, the present study addresses some relevant questions: (i) What is the effect of offsetting the normal slit from the cylinder's center on its oscillation amplitude and the frequency response? (ii) How do the slit width and reduced velocity affect the fluid-flow downstream of the slit cylinder? To address these fundamental questions, we have numerically examined the slit-cylinder with parametric variations such as the slit-offset position, slit width, and their effects on the aerodynamic loading coefficients, oscillation amplitude response, and associated flow characteristics in subsections A, B, and C of section IV.

Further, VIV-based energy harvesting devices present a potential substitute for chemical batteries as power sources for Micro-Electro-Mechanical Systems (MEMS) such as wireless sensor nodes[26, 27]. Recently, Soti et al.[28-30] comprehensively analyzed a magnet-coil type energy harvester based on electromagnetic induction. They demonstrated that the maximum average output power of such a device is independent of the length and radius of the coil and strongly depends on the cylinder's oscillation amplitude. The amplitude of oscillation is highly dependent on wind speed and direction. Increasing the operating wind speed is crucial to improve their applicability, which can be accomplished by modifying the cylinder's geometry and composition. This paper attempts to achieve this objective by the offset slit. In subsection D of section IV of this article, the effects of various slit-offset angles on the average power extracted from the oscillations of the slit-cylinder are investigated.

In addition, the energy transfer ratio (defined as the ratio of the average extracted power to the available fluid power) is reported for various offset-slit cases to demonstrate the efficacy of slit cylinders in energy harvesting.

## II. PROBLEM STATEMENT

The present study examines the effect of offsetting a normal slit from the center of an elastically mounted circular cylinder at a low Reynolds number of 150. The two-dimensional equations are solved numerically over the computational domain shown in Fig.1 (a). The computational domain spans between $-20 \leq x/D \leq 30$ and $-20 \leq y/D \leq 20$, with the cylinder's center located at the origin. The cylinder diameter is D, the slit width is s, and the $\alpha$ denotes the slit-offset position in terms of the slit-offset-angle, which is calculated as the angle between the slit center line and the line passing from the cylinder's center intersecting the slit-centerline. The slit-inclination angle (the angle between the slit center line and the cylinder's centerline) $\pi/2$ is taken throughout the paper. Fig. 1(b) shows the cross-sectional view of the slit-cylinder and defines the slit-offset-angle. The cylinder rests on the elastic supports with the help of a spring-damping arrangement, allowing it to vibrate in the transverse direction only. The mass ratio ($m^*$) is taken as 10, and the damping ratio of the system is 0.02. The incoming flow is assumed to be uniform and steady, while the advective boundary condition is applied at the outlet. The top and bottom walls are treated with the freestream boundary condition, and the no-slip condition is applied to the cylinder surface.

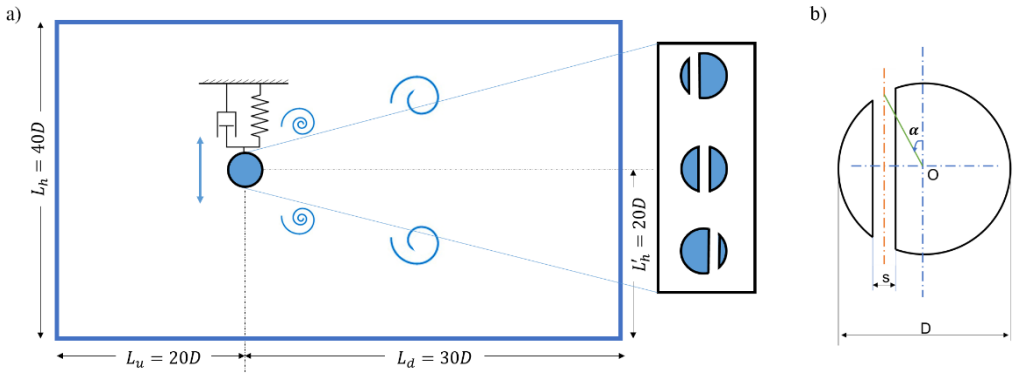

**FIG. 1.** (a) Schematic of the computational domain for the study, (b) definition of the slit-offset angle

## III. NUMERICAL DETAILS

### A. Flow Solver

The flow is assumed to be incompressible, laminar, and viscous. To model the fluid flow, we solve the conservation equations for mass and momentum as follows:

$$\nabla \cdot \vec{v} = 0 \tag{1}$$

$$\rho\left[\frac{\partial \vec{v}}{\partial t} + (\vec{v} \cdot \nabla)\vec{v}\right] = -\nabla p + \mu \nabla^2 \vec{v} \tag{2}$$

Where $\vec{v}$ represents the velocity vector, $\rho$ is the fluid density, $p$ is the static pressure, and $\mu$ is the fluid's dynamic viscosity. The Reynolds number is defined based on the cylinder diameter and the kinematic viscosity ($\nu$) as $\text{Re} = \frac{UD}{\nu}$. The study is performed at a Reynolds number of 150 to assess the effect of the slit offset on the VIV characteristics. An open-source CFD solver, OpenFOAM[31], is used to solve the flow conservation equations. The pressure-velocity coupling is felicitated using the PIMPLE (Pressure Implicit Method for Pressure Linked Equations) algorithm. Second-order discretization schemes are invoked to address all the spatial and temporal terms in the governing equations.

**B. Structural Solver**

The elastically mounted cylinder is attached to a magnet that can move along the axis of an electrically conducting coil. According to Faraday's law of electromagnetic induction, the motion of the magnet produces an electromagnetic force across the coil. The motion of the cylinder-magnet assembly can be numerically modeled as:

$$m\frac{d^2 Y}{dt^2} + 2m\zeta_s \omega_n \frac{dY}{dt} + kY = F_Y + F_m \tag{3}$$

Where $m$ is the mass of the cylinder-magnet assembly, $\zeta_s$ is the structural damping ratio, $\omega_n$ is the natural frequency of the system, k is the spring constant, $F_Y$ is the lift force acting on the cylinder surface. If the coil is connected to a resistive load, then the current is induced in the coil. The induced current opposes the motion of the magnet by applying an electromagnetic force ($F_m$), given by the following expression

$$F_m = -2m\zeta_m \omega_n \frac{dY}{dt} \tag{4}$$

Where $\zeta_m$ represents the electromagnetic damping constant. Using the above definition, Eq. (3) becomes,

$$m\frac{d^2Y}{dt^2} + 2m(\zeta_s + \zeta_m)\omega_n \frac{dY}{dt} + kY = F_Y \quad (5)$$

$$m\frac{d^2Y}{dt^2} + 2m\zeta\omega_n \frac{dY}{dt} + kY = F_Y \quad (6)$$

The total damping coefficient $\zeta$ in Eq. 6 includes the damping coefficient due to losses in the transmission system ($\zeta_s$) and the added damping coefficient ($\zeta_m$) for electromagnetic energy harvesting. $m$ represents the total mass of the system (mass of the cylinder and added mass), $\omega_n$ represents the natural frequency of the system ($\omega_n = \sqrt{\frac{k}{m}}$), $\zeta$ is the damping coefficient ($\zeta = \frac{c}{2\sqrt{km}}$), and $k$ is the spring constant. The total fluid force in the transverse direction ($F_Y$) is obtained by integrating the pressure and viscous forces acting on the cylinder's surface in the transverse direction. Following the previous literature on VIV, the motion of such a system can be represented using some of the non-dimensional parameters such as non-dimensional mass-ratio $\left(m^* = \frac{m}{\frac{\pi}{4}\rho D^2 L}\right)$, non-dimensional spring constant $\left(k^* = \frac{k}{\rho U_\infty^2 L}\right)$, non-dimensional damping coefficient $\left(C^* = \frac{c}{\rho U_\infty DL}\right)$, non-dimensional reduced velocity $\left(U_r = \frac{U_\infty}{f_n D}\right)$, non-dimensional frequency $\left(f = \frac{f_s D}{U}\right)$, non-dimensional time $\left(\tau = \frac{tU}{D}\right)$, and non-dimensional mean transverse oscillation amplitude $\left(A_y^* = \frac{(Y/D)_{max} + abs(Y/D)_{min}}{2}\right)$.

## C. Fluid-Structure Coupling and Mesh Motion

Equations (1), (2), and (6) need to be solved in a coupled manner to obtain accurate simulation data. Fluid force is calculated by solving Eqn. (1) and (2) in each timestep, the force is fed to Eqn. (6) to obtain the updated cell locations. The weakly coupled form of the structural equation is implemented as the explicit function object in the OpenFOAM framework, in which the flow and structural equations are solved independently and sequentially, with coupling invoked by forces and boundary conditions. The algorithm is described in detail below.

1. Determine the updated pressure and velocity fields by advancing the Fluid solver in time.

2. Calculate the lift and drag forces on the cylinder's surface based on the current pressure and velocity utilizing the *libforces* library of OpenFOAM.

3. Use these lift forces as the external forces applied to the structural equations and integrate them using the fourth-order Runge Kutta method to determine the updated position and velocity of the cylinder at each time step.

4. Perform the mesh deformation based on the motion of the boundaries utilizing the dynamicMeshDict library of OpenFOAM.

The loosely coupled methods are extensively used in the literature[32-34]. The detailed implementation of the weakly coupled form of the structural equation can be found in work by Jester & Kallinderis[35] and Carmo et al.[36]. The timestep of the fluid solver is kept sufficiently small for better convergence for the coupling. To incorporate the mesh-motion calculated from the Eqn. (6) the positions of the finite volume cells are computed by the Laplace equation,

$$\nabla \cdot (\gamma_m \nabla z) = 0 \qquad (7)$$

Where $\gamma_m$ is the mesh-diffusion coefficient, and z is the cell displacement field. Due to the channel's fixed top and bottom boundaries, the mesh motion is distributed through the grid using the inverse mesh diffusion method. The mesh diffusion is calculated based on the inverse distance from the cylinder body[37, 38]. The diffusivity field is based on the quadratic relation to the inverse of the cell center distance ($l$) to the nearest boundary, i.e., $1/l^2$.

### D. Energy Harvesting and Power Extraction

The VIV oscillations of the slit-cylinder can be efficiently utilized to extract power by attaching a magnet with the cylinder along the axis of the conducting wire coil. This arrangement constitutes a small electric generator unit based on electromagnetism. As the magnet moves periodically across the coil, it induces an electric current in the conducting coil. These eddy currents also apply a drag force on the moving magnet (Lenz's law), which can be numerically modeled as the electromagnetic damping term used previously by Soti et al.[28, 29]. They demonstrated that the power extracted from the 1-DOF VIV system is equal to the power dissipated by the damper, and numerically it can be given as,

$$P(t) = \frac{cv_y^2}{\frac{1}{2}\rho U^3 DL} = \frac{c\dot{y}*^2}{\frac{1}{2}\rho U^3 DL} \qquad (8)$$

Where P(t) represents the instantaneous power extracted. The average extracted power over an oscillation cycle can be estimated by assuming the sinusoidal and periodic form of the cylinder oscillations as,

$$P_{avg} = \frac{1}{T}\int_0^T P(t)dt \qquad (9)$$

Where T is the period of oscillation. For the steady-state, large-amplitude vortex-induced oscillations, the fluid force and the body displacement response oscillate at the same frequency (as seen in FIG. 7 (c)), and the cylinder displacement y* can be represented as $y^* = A_Y^* \sin(2\pi f t)$. Using this, the average extracted power is equal to,

$$P_{avg} = 4\pi^4 m^* \zeta \frac{((f_s/f_n)A_Y^*)^2}{U_r^3} \qquad (10)$$

Here $P_{avg}$ represents the non-dimensional average extracted power. This energy of the oscillating cylinder is originally extracted from the kinetic energy of the flow for the projected area of the cylinder (projected area: (D+2A)L), which can be given as,

$$P_{fluid} = \frac{1}{2}\rho U^3 (D+2A_Y)L = \frac{1}{2}\rho U^3 DL(1+2A_Y^*) \qquad (11)$$

Thus, the energy transfer ratio, $\eta$, can be written as,

$$\eta = \frac{P_{avg}}{(P_{fluid}/\frac{1}{2}\rho U^3 DL)} = \frac{P_{avg}}{(1+2A_Y^*)} \qquad (12)$$

**E. Mesh and Time-step Independence Study and Error Analysis**

As the study deals with the slit-cylinder, we have performed the grid-independence study over the transversely vibrating cylinder with a normal slit of slit-width (s/D) of 0.20 with a slit-offset-angle of $0^o$ at Re = 150. The mass ratio is 10, with the damping ratio and the reduced velocity as 0.02 and 5, respectively. ANSYS ICEM-CFD[39] is used to generate the mesh over the domain. O-grid blocking is used to map the cylinder surface in the near-cylinder region, as shown in Fig. 2. The $y^+$ value at the cylinder surface is kept below 0.8 with the cell expansion ratio of 1.02 in the small square section around the cylinder to resolve the wall stresses correctly. The rest of the computational domain is filled with quadrilateral cells with an expansion ratio of 1.2 to save the computational cost.

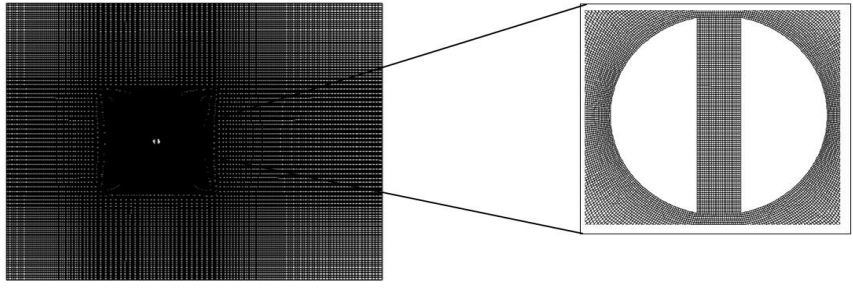

**FIG. 2.** Schematics of the computational mesh used for the study

The grid independence study is conducted on four sets of grids, i.e., Grid-1 (31,398 cells), Grid-2 (57,864 cells), Grid-3 (90,978 cells), and Grid-4 (143,784 cells). The temporal variation of the aerodynamic load coefficient, i.e., $C_L$ over the cylinder, and the corresponding transverse oscillation amplitude (normalized with the cylinder diameter, D) for different grids and their variation with the mesh size is shown in Figure 3 (a) and (b), respectively. We observed that the differences between Grid-3 and Grid-4 drop to less than 1% (as seen in Table-1). Hence, Grid-3 seems sufficient to carry out this study.

Further, to develop more confidence in the grid, we have also looked at the Grid-Convergence Index (GCI) proposed by Roache[43, 44] and extensively used in past literature. A detailed description of the GCI calculations is given in Appendix-A. We have considered the r.m.s. value of the lift coefficient ($C_{L_{rms}}$) for the GCI study. The factor of safety for best estimation is taken as 1.25. The results are duly tabulated in Table II. We observe the reduction of both the parameters for the constitutive grid refinements. The calculations confirm that the Grid-3 is nicely resolved. Fig. 3 (c) reports the order of accuracy in terms of the $L_2$ norm of the errors between the grids against the grid spacing (h) on a log-log scale. The slope of the $L_2$ norm is compared with the theoretical slope of order 2. The order of accuracy is found to be 1.928. This value is used to calculate the GCI for different grids.

**TABLE I.** Grid independence study for the moving cylinder [Bold values represent the values for the mesh selected for the current study]

| Grid | No. of Cells | $C_{L_{rms}}$ | % Change in $C_{L_{rms}}$ | $A_Y^*$ | % Change in $A_Y^*$ |
|---|---|---|---|---|---|
| Grid-1 | 31,598 | 0.6228 | - | 0.4250 | - |
| Grid-2 | 57,864 | 0.6879 | 9.46 % | 0.4254 | 0.09 % |
| **Grid-3** | **90,978** | **0.7081** | **2.85 %** | **0.4258** | **0.09 %** |
| Grid-4 | 143,784 | 0.7089 | 0.11 % | 0.4258 | 0.00 % |

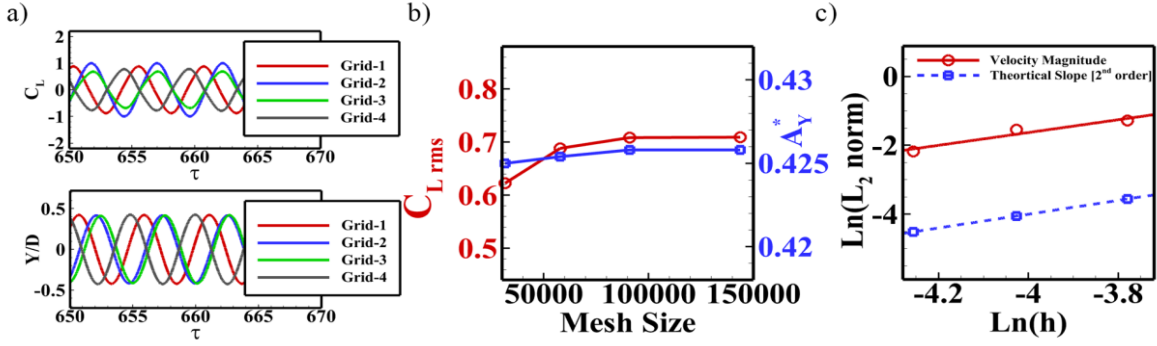

**FIG. 3.** Grid Independence study at Re = 150 for an unconfined cylinder with a normal slit: (a) time history of aerodynamic lift coefficient ($C_L$) and corresponding transverse oscillation amplitude ($Y/D$), (b) $C_{L_{rms}}$ variation with the mesh size, and (c) $L_2$ norm against the grid spacing (h) compared with the theoretical slope for $2^{nd}$ order [ $m^* = 10, \zeta = 0.02, U_r = 5, \alpha = 0^o, s/D = 0.20$, Re = 150 ]

**TABLE II.** Richardson error estimation and grid-convergence index for three sets of grids

|  | $r_{CB}$ | $r_{DC}$ | o | $\varepsilon_{CB}(\times 10^{-2})$ | $\varepsilon_{DC}(\times 10^{-3})$ | $E_2^{coarse}$ | $E_1^{fine}$ | $GCI^{coarse}$ | $GCI^{fine}$ |
|---|---|---|---|---|---|---|---|---|---|
| $C_{L\,rms}$ | 1.25 | 1.25 | 1.928 | 2.936 | 1.130 | 0.084 | 0.002 | 10.498 % | 0.263 % |

After selecting Grid-3 based on the grid-independence study, we have further performed the time step independence study for four different time steps: 0.005, 0.001, 0.0005, and 0.0001. The results are tabulated in Table III below. The difference in the calculated values of the lift coefficient, $C_{L_{rms}}$ (obtained from the fluid solver), and the mean oscillation amplitude, $A_Y^*$ (obtained from the structural response), for time-step 5e$^{-4}$ was found to be less than 3% and 1 %, respectively, as compared to the timestep of 1e$^{-3}$. Hence, the time-step of 5e$^{-4}$ is selected to carry out the further simulations.

**TABLE III.** Time-step independence study for the moving cylinder with Grid-3 [Bold values represent the values for the mesh selected for the current study]

| Time-step value (s) | Fluid response | | Structural response | |
|---|---|---|---|---|
| | $C_{L_{rms}}$ | % Change in $C_{L_{rms}}$ | $A_Y^*$ | % Change in $A_Y^*$ |
| 5e$^{-3}$ | 0.761 | - | 0.447 | - |
| 1e$^{-3}$ | 0.693 | 8.879 | 0.428 | 4.224 |
| **5e$^{-4}$** | **0.678** | **2.133** | **0.425** | **0.606** |
| 1e$^{-4}$ | 0.671 | 1.119 | 0.423 | 0.516 |

## F. Numerical Validation

Before addressing the main problem statement, we have validated our numerical setup with the available literature on the 1-DOF VIV for a low Re of 150. Figure 4 (a) shows the computational domain used for the validation study. The mass ratio of the cylinder is 2, and the damping coefficient is taken as 0. Figure 4 (b) compares the non-dimensional mean transverse oscillation amplitude with the previously published literature. The results depict the two-branch response: the initial branch (IB, $U_r \leq 4$) and the lower branch (LB, $4 \leq U_r \leq 8$). The results are in good agreement for both the branches. The maximum value of non-dimensional amplitude (Y/D) is 0.5718 in our simulations. The same was predicted to be 0.5624, 0.5672, 0.5563, and 0.5658 by Zhao et al.[40] in their 2D and 3D simulations, Ahn & Kallinderis[41] in their experiments, and Bao et al.[42] and differs by 2% from our results. Further, the frequency response is compared with the available data by Zhao et al.[40] in Fig. 4 (c), which also agrees well with the literature.

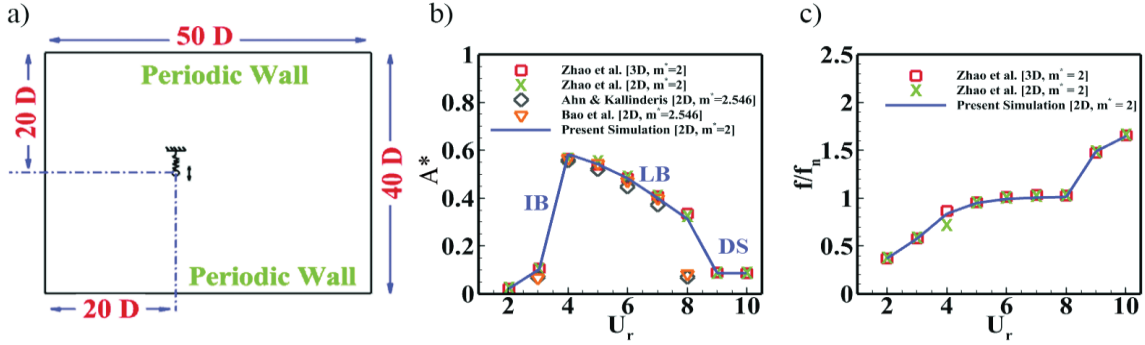

**FIG. 4.** (a) Schematic of the benchmark VIV used to validate the numerical framework, (b) comparison of the non-dimensional oscillation amplitude ($A^*$) with the published results over the range of reduced velocity ($U_r$), and (c) comparison of the frequency response ($f/f_n$) [$m^* = 2, \zeta = 0$, Re = 150]

## IV. RESULTS AND DISCUSSION

We have used flow visualization techniques to capture and understand the fluid dynamic behavior associated with the different offset slits. The instantaneous vorticity contours, streamline patterns, and the aerodynamic lift force for different slit-offset angles compared to the cylinder with no slit case. The amplitude and frequency response are analyzed to identify the VIV response branches for the slit-offset cases. Further, the effect of different geometrical and VIV parameters such as slit width, reduced velocities, slit-offset angles, etc., is investigated to assess the VIV characteristics. The last section details the implications of the offset slits in energy harvesting applications.

### A. Effect of slit-width

To confirm the effect of slit width on VIV features, we simulated the case of a normal slit with a slit-offset angle of $0^o$, using various slit widths ranging from $0.10 \leq s/D \leq 0.30$ at Re = 150. The s/D = 0 case represents the case of the cylinder with no-slit. The mass and damping ratio is 10 and 0.02, respectively. The simulations are conducted at a reduced velocity of 5 since they fall inside the no-slit case's lock-in regime. Figure 5 depicts the fluid-dynamic properties of the flow in terms of the instantaneous z-vorticity distribution, time-averaged velocity streamlines over the cylinder with varying slit widths, the lift coefficient, and the transverse oscillation amplitude. The wake behind the slit cases with the smaller slit widths ($s/D \leq 0.15$) resembles the no-slit case, as seen in Fig. 5(a). The slit cases exhibit one pair of counter-rotating vortices at the slit openings for this range of slit width (see Fig. 5(b)), which supports the alternate suction and blowing of the main flow into the slit. The time history of the lift-force coefficient and the transverse oscillation amplitude (normalized with the cylinder diameter, D) is shown in Fig. 5(c). There is a minor phase discrepancy between the lift-force and the oscillation amplitude at smaller slit widths, decreasing to zero as the slit width increases.

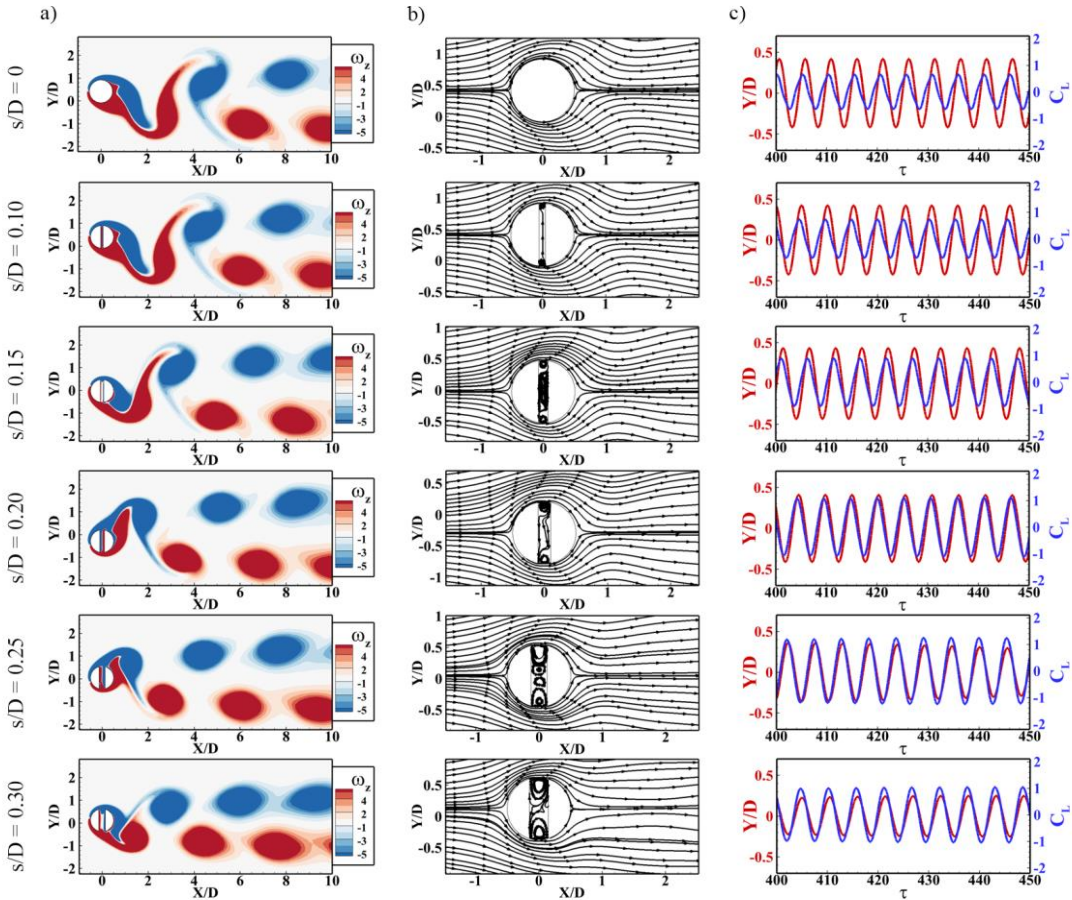

**FIG. 5.** Flow characteristics behind the moving cylinder for different slit widths: (a) instantaneous z-vorticity ($\omega_z$) contour, (b) time-averaged velocity streamlines, and (c) temporal evolution of non-dimensional transverse oscillation amplitude ($Y/D$) and the corresponding lift-coefficient ($C_L$) [ $m^* = 10, \zeta = 0.02, U_r = 5, \alpha = 0^o$, Re = 150 ]

Fig. 6 (a) reports the non-dimensional mean transverse oscillation amplitude ($A^*$) and the corresponding frequency response in terms of the Strouhal no. (St). For the low slit widths ($s/D \leq 0.15$), the oscillation-amplitude and frequency response remain almost constant over the slit width. The lift and the drag force on the cylinder surface continuously increase till $s/D \leq 0.20$ due to the additional flow into the slit from the main flow, as seen in Fig. 6(b). At $s/D = 0.25$, the mean-streamline shows the two pairs of the counter-rotating vortices inside the slit. Also, the vortices near the slit openings grow in size. This results in the reduced pressure acting on the surface, and the lift and drag coefficient drops to lower values. For higher slit widths ($s/D \geq 0.25$), the shed vortices travel downstream closer, resulting in a stabilized wake and reduced cylinder oscillations with a lower frequency. Based on the preceding discussions, we have used $s/D = 0.20$ for the subsequent simulations.

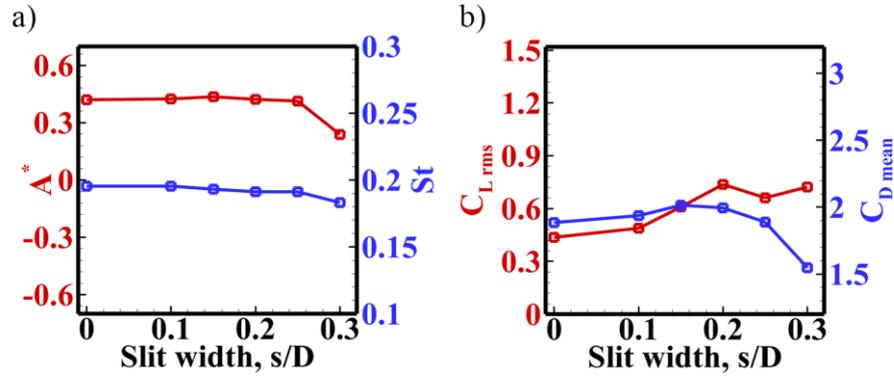

**FIG. 6.** Effect of slit width ($s/D$) on the VIV characteristics and aerodynamic performance: (a) non-dimensional mean transverse oscillation amplitude ($Y/D$) with the corresponding Strouhal no. (St), and (b) r.m.s. value of the aerodynamic lift coefficient ($C_{L_{rms}}$) and mean drag coefficient ($C_{D_{mean}}$) [ $m^* = 10, \zeta = 0.02, U_r = 5, \alpha = 0^o$, Re = 150 ]

## B. Effect of Slit Offset

The preceding section emphasizes the effect of varying the slit width for an elastically mounted slit cylinder with a normal slit passing symmetrically through its centerline. As a result, the slit-offset-angle ($\alpha$), defined as the angle between the line passing through the center of the slit cylinder and the slit's centerline, was held constant. The main objective of this section is to see how the flow behaves when the normal slit is shifted to either the back stagnation point (corresponding to the negative values $\alpha$) or the front stagnation point (corresponding to the positive values $\alpha$). As the flow passes over the cylinder, the pressure decreases due to velocity increase, resulting in the thickening of the boundary layer. Based on the pressure difference between the top and bottom of the cylinder, the slit-vent alternatively draws the main flow from the boundary layer and blows it out from the other end, termed suction and blowing in literature[10, 23, 24]. Fig. 7(a) portrays the time-averaged velocity streamlines over the slit-cylinder for different slit offset angles. The slit at the cylinder's center ($\alpha = 0^o$) exhibits the one pair of counter-rotating vortices placed at the slit-openings. While offsetting the slit towards the back stagnation point (for $\alpha < 0^o$), fills the slit with two pairs of counter-rotating vortices (refer to Fig. 7(a)). From the instantaneous vorticity contour in Fig. 7(b), we observe that there is not much effect on the vortex shedding behind the cylinder on offsetting the slit towards the back-stagnation point. The lift force and the oscillation amplitude show a slight phase shift compared to the $\alpha = 0^o$ case (refer to Fig. 7(c)). Also, for $\alpha = -30^o$ the case, the slit-cylinder sheds vortices a little closer.

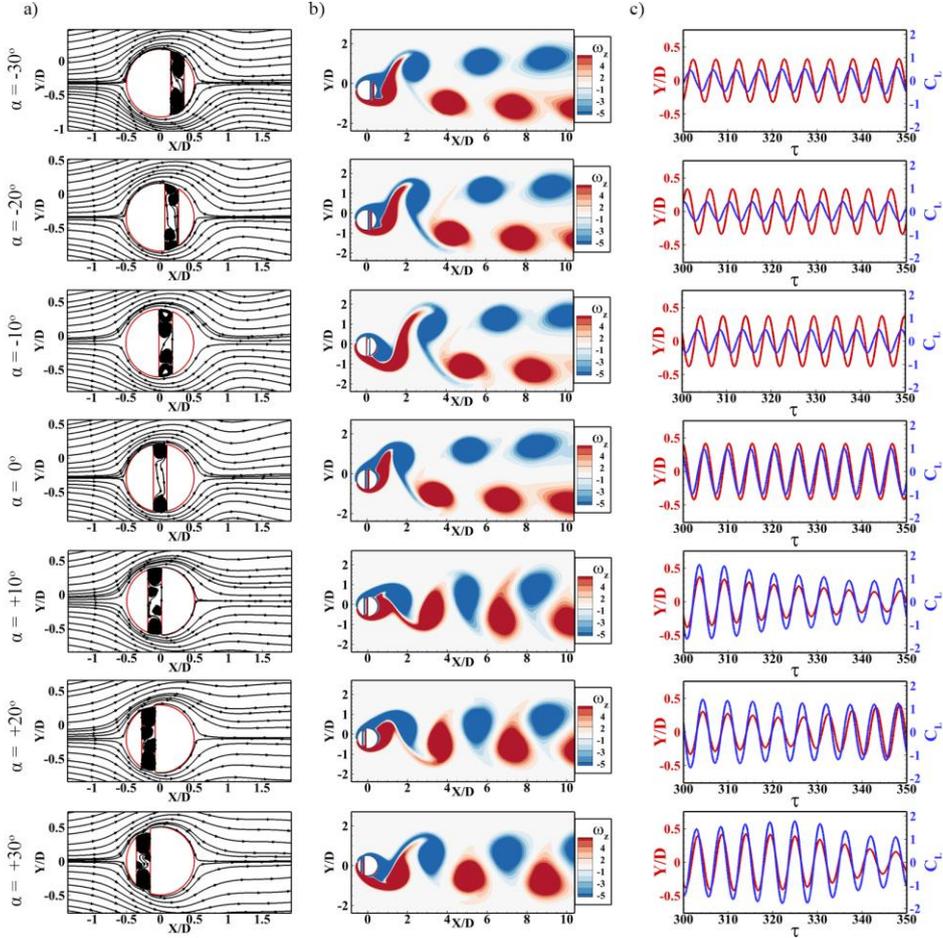

**FIG. 7.** Effect of offsetting the normal slit from the center: (a) time-averaged velocity streamlines, (b) instantaneous z-vorticity ($\omega_z$) contour behind the cylinder, and (c) time history of the non-dimensional oscillation amplitude ($Y/D$) and corresponding lift coefficient ($C_L$) [ $m^* = 10, \zeta = 0.02, U_r = 5, s/D = 0.20$, Re = 150 ] ($\alpha$ represents the slit-offset angle)

Fig. 8 (a) shows the non-dimensional mean transverse oscillation amplitude and the corresponding oscillation frequency for different slit-offset-angles. We see that offsetting the slit towards the negative values of the slit-offset-angle results in reduced oscillation amplitude and increased oscillation frequency. The reduction in the oscillation amplitude is accompanied by the lower values of the aerodynamic lift and the drag (as seen in Fig. 8 (b)). When the slit is shifted towards the front stagnation point, the vortex shedding changes to the standard Von-Karman street type of shedding. The cylinder alternatively sheds the vortices along the wake centerline (refer to Fig. 7(b)). The lift and the oscillation amplitude are in phase, and the cylinder oscillates with the higher oscillation amplitude and lower oscillation frequency. This is accompanied by the increased lift force and the reduced drag force over the cylinder surface.

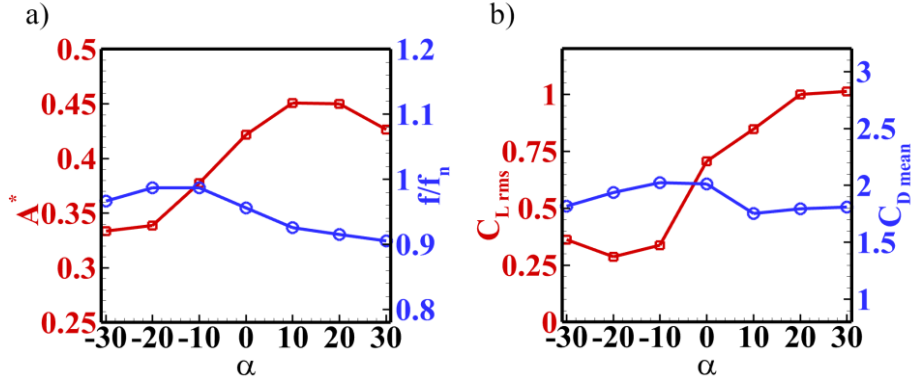

**FIG. 8.** (a) The non-dimensional mean transverse oscillation amplitude ($A^*$) and the corresponding oscillation frequency normalized with the natural frequency ($f/f_n$), (b) r.m.s. value of the aerodynamic lift coefficient ($C_{L_{rms}}$) and the time-averaged mean drag coefficient ($C_{D_{mean}}$) [$m^* = 10, \zeta = 0.02, U_r = 5, s/D = 0.20$, Re = 150] ($\alpha$ represents the slit-offset angle)

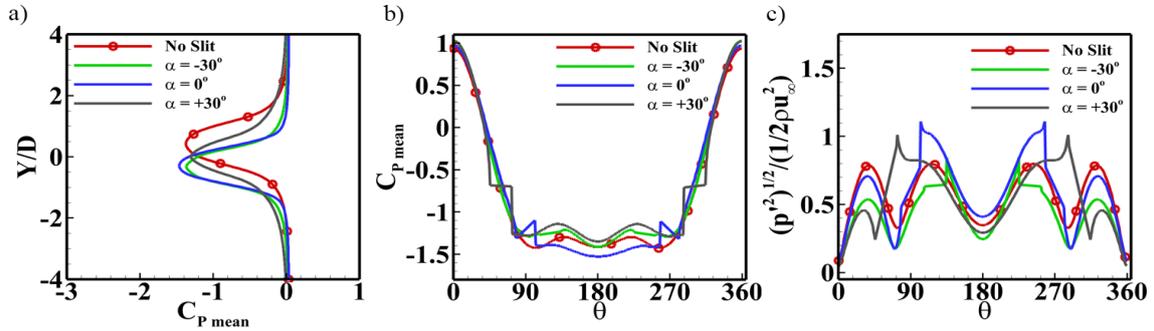

**FIG. 9.** Mean Pressure coefficient ($C_{P_{mean}}$) distribution for different slit-offset-angles ($\alpha$): (a) in the wake at x/D = 0.6, (b) over the cylinder surface; and (c) mean pressure fluctuation over the cylinder surface ($\theta$ represents the azimuthal angle)

Further, Fig. 9 (a) illustrates the mean pressure coefficient behind the cylinder in the wake at x/D = 0.6. The mean pressure distribution is the same for all the cases, although the pressure recovery is maximum in the wake for $\alpha = +30^o$ case. Other cases exhibit almost the same mean pressure value as the no-slit case in the wake. Fig. 9 (b) depicts the distribution of the mean pressure coefficient over the cylinder surface and is duly compared to the no-slit case. We observe the higher pressure recovery after the slit on the back side of the cylinder for $\alpha = +30^o$ case, which results in the higher lift coefficient values for these cases. For $\alpha = 0^o$ case, the pressure follows the distribution on the cylinder surface similar to the no-slit case before the slit; just after the slit, the boundary layer separates, and the pressure decreases much further, as seen in Fig. 9 (b). Fig. 9 (c) displays the pressure fluctuations over the cylinder surface for different slit-offsets. For the no-slit case, the maximum pressure fluctuations occur near the separation point on the front surface, and the minimum pressure fluctuations are at the rear stagnation point $\theta = 180^o$. The second maximum peak around $\theta = 126^o$ denotes the separation of the reattached flow over the surface. We observe that the slit cases also follow a similar trend with lower peaks for the pressure fluctuations. Also, the separation point shifts upstream for $\alpha = +30^o$ case. The pressure fluctuations increase drastically after the slit along the cylinder's surface, taking the minimum value at the rear stagnation point. On the lower side of the cylinder, the flow behaves in the same manner.

## C. Effect of Reduced Velocities and VIV Response Branches

Looking at the effects of different offset slits, we find that offsetting the slit towards the front stagnation point results in improved transverse oscillations, which makes it suitable for energy harvesting applications. Offsetting it towards the back stagnation point results in suppressed oscillations. Now to confirm this behavior over the reduced velocities, this section deals with the VIV response of the two above-mentioned slit cases ( $\alpha = -30^o$ and $\alpha = +30^o$ ) and compares it with the response of the cylinder with no slit over the range of reduced velocities ( $3 \leq U_r \leq 8$ ).

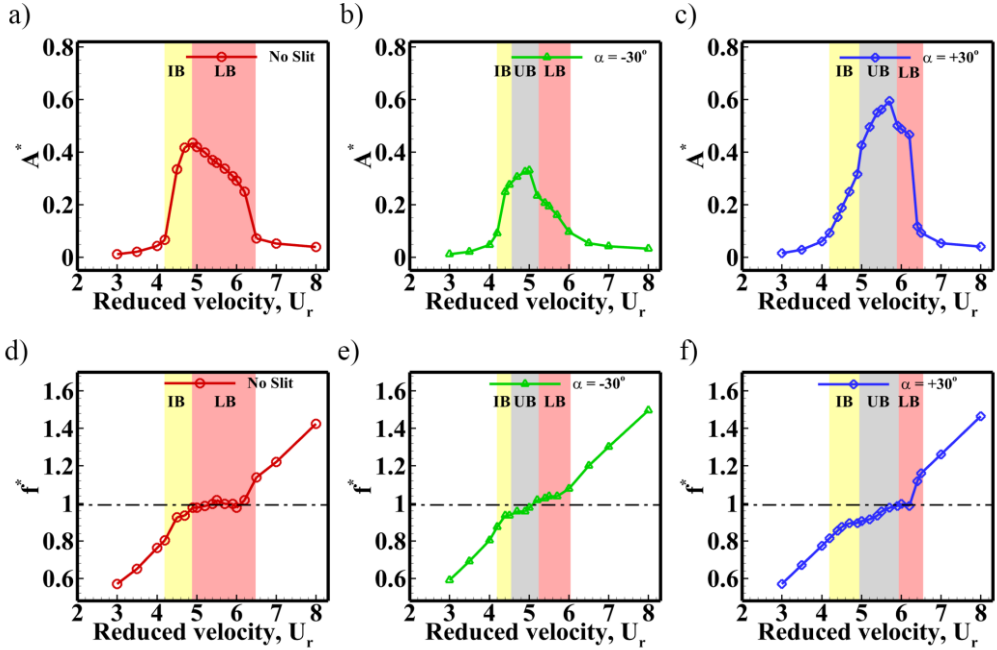

**FIG. 10.** VIV response for different slit-offset angles at varying reduced velocity ($U_r$): (a), (b), and (c) depicts the non-dimensional mean oscillation amplitude ($A^*$); and (d), (e), and (f) depict the response of oscillation frequency normalized by the natural frequency of the system ($f^* = f / f_n$) [ $m^* = 10, \zeta = 0.02, s/D = 0.20$ , Re $= 150$ ]

Figure 10 (a), (b), and (c) report the mean-transverse oscillation amplitude for three different cases (no-slit case, $\alpha = -30^o$ case, $\alpha = +30^o$ case) and their corresponding frequency response over the range of reduced velocities in Fig. 10 (d), (e) and (f), respectively. The case of the no-slit exhibits the two-branch response, i.e., Initial Branch (IB, $4.2 \leq U_r \leq 4.9$ ) and Lower Branch (LB, $4.9 < U_r \leq 6.3$ ). In IB, the amplitude linearly increases with the increase in the reduced velocity and attains the maximum oscillation amplitude value. The frequency during this branch deviates from the standard Strouhal frequency slope. Fig. 11 (a) shows the variation of the aerodynamic coefficients (i.e., r.m.s. value of the lift coefficient and the time-averaged drag coefficient) for the no-slit case over the range of reduced velocities. We observe that the lift and drag continuously increase in the Initial branch causing the linear increase in the oscillation amplitude. On increasing the reduced velocity further, the oscillation amplitude, aerodynamic lift, and drag coefficients start to fall. During LB, the frequency response remains almost constant and returns to the linear slope of the Strouhal frequency. This is in agreement with the previous studies on 1-DOF VIV[4,5]. For $\alpha = -30°$ case, the oscillation amplitude response exhibits the three-branch response, i.e., Initial Branch (IB, $4.2 \leq U_r \leq 4.6$), Upper Branch (UB, $4.6 < U_r \leq 5.2$), and Lower Branch (LB, $5.2 < U_r \leq 6$). The initial branch starts when the oscillation amplitude increases linearly with the reduced velocity due to the increase in the lift coefficient (as can be seen in Fig. 11 (b)), and the frequency leaves the linear slope of the Strouhal no. The upper branch shows the maximum oscillation amplitude against the continuous decaying lift and drag coefficients over the range of the reduced velocities. The frequency lock-in occurs lower than the system's natural frequency. This maximum oscillation amplitude and the r.m.s. value of the lift coefficient is less than the no-slit case due to the placement of the slit near the back-stagnation point. In the lower branch, the oscillation amplitude starts to decay with a constant lift coefficient and the decaying drag coefficient, and the frequency lock-in is observed at a higher oscillation frequency than the natural frequency for some reduced velocities, which returns to the linear dependence at the end of the lower branch. After this branch, the lift and drag almost become constant, and the cylinder oscillates with very low-amplitude oscillations.

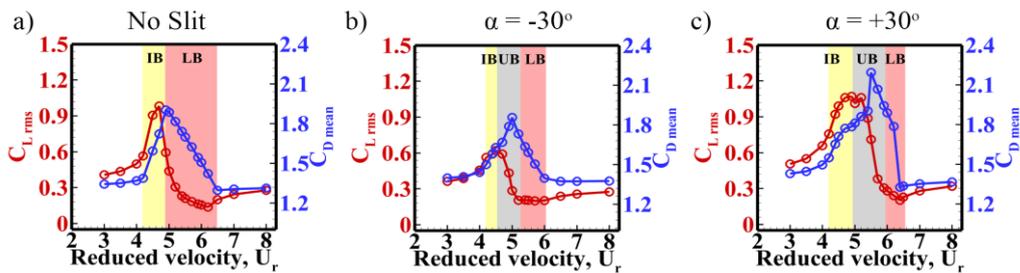

**FIG. 11.** Aerodynamic performance of the slit cylinder for different slit offsets at varying reduced velocities (in terms of the r.m.s. value of the aerodynamic lift coefficient ($C_{L_{rms}}$) and the time-averaged mean drag coefficient ($C_{D_{mean}}$)): (a) for no-slit case, (b) for $\alpha = -30°$ case, and (c) for $\alpha = +30°$ case [$m^* = 10, \zeta = 0.02, s/D = 0.20$, Re = 150] ($\alpha$ represents the slit-offset angle)

Further, Fig. 10 (c) and (f) show the amplitude and the frequency response for $\alpha = +30°$ case. It also exhibits the three-branch response, i.e., the initial branch (IB, $4.2 \leq U_r \leq 5$), the upper branch (UB, $5 < U_r \leq 6$), and the lower branch (LB, $6 < U_r \leq 6.5$). During the initial branch, the oscillation amplitude and the lift and drag coefficients (as reported in Fig. 11 (c)) increase linearly with the increase in the reduced velocity. The frequency also varies linearly during this branch of VIV response. In the upper branch, the oscillation amplitude continuously increases with almost constant lift-coefficient reaching the maximum oscillation amplitude value and starts to decay slowly with the decrease in the lift-coefficient value. The frequency increases linearly with the reduced velocity exhibiting the lock-in response at the system's natural frequency. The oscillation amplitude and the lift and drag coefficients decrease rapidly in the lower branch. The oscillation frequency varies with a linear slope higher than the Strouhal frequency for the lower branch and returns to the Strouhal frequency slope at the end of the lower branch. The lift and drag become almost constant, and the cylinder oscillates with very low amplitude.

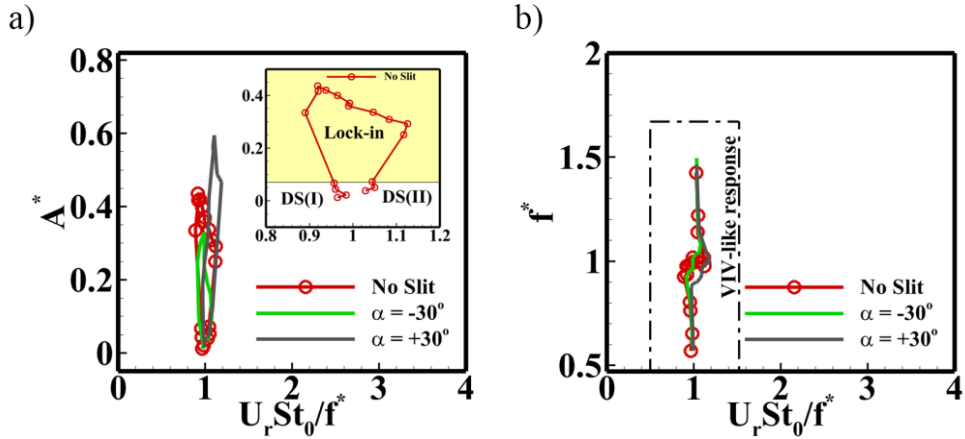

**FIG. 12.** Flow over the cylinder with or without the offset-slits undergoing VIV: (a) non-dimensional mean oscillation amplitude of cylinder vibration ($A^*$), (b) frequency response relative to the non-dimensional parameter $U_r St_0 / f^*$, where $St_0 = \dfrac{f_{s_0} D}{U}$ is the Strouhal number based on the vortex shedding frequency of the stationary cylinder ($f_{s_0}$), and represents the oscillation frequency normalized with the natural frequency of the system ($f^* = f / f_n$) [The zoomed view of the amplitude curve for VIV-like response of no-slit case is shown as an inset in (a)]

Figure 12 (a) and (b) report the oscillation amplitude and frequency response, respectively, over the cylinders with/without slit concerning the non-dimensional parameter, $U_r St_0 / f^*$, where $St_0$ is the Strouhal no. corresponds to the vortex shedding frequency for the flow past the stationary (slit) cylinder. This plot confirms if the oscillation response belongs to VIV or if it achieves galloping. The no-slit case's amplitude and frequency response curve exhibit the VIV-like characteristics[43]. Similar behavior is observed for the offset-slit cylinder cases, although the amplitude curve for the slit cases is more skewed. The zoomed view of the amplitude curve shows the smooth transition between the desynchronization and lock-in for the no-slit case.

## D. Implications in the Energy Harvesting and Power Extraction

Due to the worldwide increasing energy demand, there has been a quest to find new ways of harvesting clean, renewable, and economical energy. Vortex-induced vibrations of a circular cylinder offer one of such methods for energy harvesting applications. Utilization of the oscillatory motion via electromagnetism dates back to the era of Nikola Tesla's inventions. Recently, Soti et al.[28-30] demonstrated the capability of harnessing the electrical power from the VIV motion of the circular cylinder at a low Reynolds number via a coil-magnet arrangement. In their experimental and numerical study, they attached a magnet to the cylinder axis and placed the cylinder inside a conducting wire coil. The transversely oscillating cylinder induces the electric current in the coil. These eddy currents also exert a drag force back on the cylinder and are modeled as the additional damping.

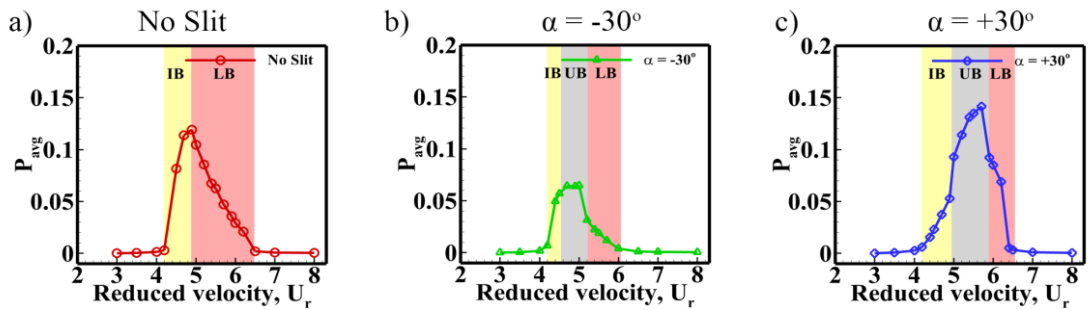

**FIG. 13.** Effect of slit-offset-angle ($\alpha$) on the harnessed power in terms of non-dimensional average extracted power ($P_{avg}$): (a) for no-slit case, (b) for $\alpha = -30^o$ case, and (c) for $\alpha = +30^o$ case [ $m^* = 10, \zeta = 0.02, s/D = 0.20$, Re = 150 ]

Figure 13 provides the non-dimensional average power extracted from the cylinder's oscillations based on the Eqn. 10 for two different offset-slit-angles compared to the no-slit case. For the no-slit case, the power extracted increases with the reduced velocity in the Initial Branch (IB), reaching the maximum power extracted at the end of the IB. After IB, the extracted power decreases with the reduced velocity due to lower oscillation amplitude in LB. For the offset-slit cases, a similar trend is observed for both the branches (IB and LB). For $\alpha = -30^o$ case, in the Upper Branch (UB), although the oscillation amplitude continuously increases, the power extracted remains almost constant due to increasing reduced velocity (in the denominator of Eqn.10). Although, due to reduced oscillation amplitude $\alpha = -30^o$, the peak extracted power is almost half compared to the no-slit case. Thus, offsetting the slit towards the back stagnation point is not recommended for energy harvesting applications. For $\alpha = +30^o$ case, the extracted power continuously increases in the Upper Branch (UB) and achieves the highest peak power compared to the other two cases.

Figure 14 reports the energy transfer ratio (or the power extraction efficiency) for the three cases, i.e., no-slit $\alpha = -30^o$, and $\alpha = +30^o$ case. The energy transfer ratio follows the same trend as the extracted average power for all the cases. The IB exhibits the increasing efficiency with the reduced velocity, UB shows the constant (for $\alpha = -30^o$ case) or the increasing (for $\alpha = +30^o$ case) power extraction efficiency, and the LB shows the decreasing trend with the increasing reduced velocity. Although the $\alpha = +30^o$ case extracts the more considerable average power (as seen in Fig. 13), the peak energy transfer ratio is almost the same as the no-slit case. But, because of the three-branch response in $\alpha = +30^o$ case, we observe the peak efficiency over the wider range of reduced velocities. This makes it a suitable and preferred case for the applications in energy harvesting via VIV, among others. As most previous studies have investigated the combined effect of the mass and damping on energy extraction, we have analyzed it independently using the non-dimensional combined parameter[46], $(A^* f^*)^2 / U_r^3$. If all other parameters are fixed, Fig. 15 reports the linear dependence of the energy transfer ratio on this combined parameter. All three cases exhibit almost the same slope of the curve, although there are different peaks due to the difference in the energy transfer ratios.

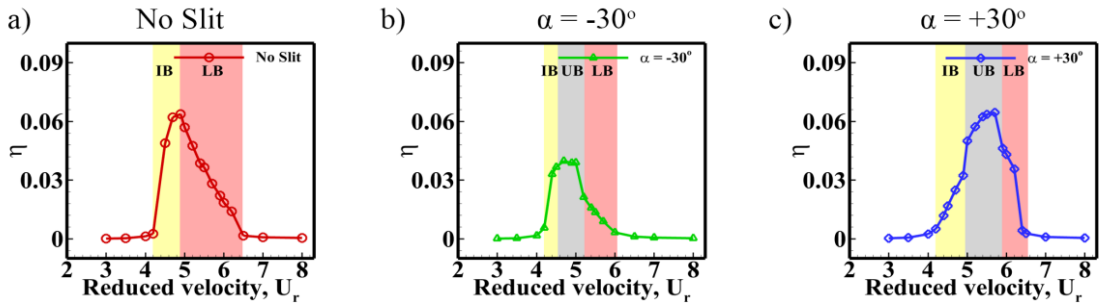

**FIG. 14.** Effect of slit-offset-angle ($\alpha$) on the energy transfer ratio (or power extraction efficiency) ($\eta$): (a) for no-slit case, (b) for $\alpha = -30^o$ case, and (c) for $\alpha = +30^o$ case [ $m^* = 10, \zeta = 0.02, s/D = 0.20$, Re $= 150$ ]

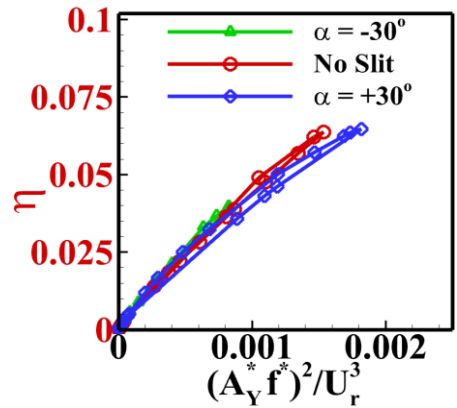

**FIG. 15.** Energy transfer ratio ($\eta$) dependence on the combined parameter, $(A^* f^*)^2 / U_r^3$ [ $m^* = 10, \zeta = 0.02, s/D = 0.20$, Re $= 150$ ]

## V. CONCLUSION

The present study investigates the role of the slit offset on VIV characteristics of an elastically mounted circular cylinder, bounded to oscillate in the transverse direction only. The effect of various parameters such as slit-width (s/D), slit-offset-angle ($\alpha$), the reduced velocities ($U_r$), and its implications in energy harvesting applications are widely investigated using the mean-transverse oscillation amplitude and frequency response. As the fluid flows over the slit-cylinder with the slit placed at the center of the cylinder, periodic blowing and suction establish, resulting in the increased lift coefficient. As we shift the slit towards the back stagnation point, the cylinder sheds alternate vortices away from each other, resulting in the reduced Reynolds stress behind the cylinder and a stabilized wake. This results in the suppression of the cylinder's oscillations. When we shift the slit towards the front stagnation point, the flow separates early from the cylinder's surface, resulting in the Von-Karman street-like shedding of alternate vortices. It results in the higher-pressure recovery after the slit on the back side of the cylinder, which increases the lift coefficient and the oscillation amplitude, making it suitable for energy harvesting applications. Further, this behavior of offset-slits is confirmed over the wide range of the reduced velocities. We observe that the slit-cylinder exhibit the three-branch VIV response (i.e., initial branch (IB), upper branch (UB), and lower branch (LB)). The upper branch is absent in the case of the no-slit cylinder, and it exhibits the two-branch VIV response (i.e., IB and LB) at the low Reynolds number of 150. Further, to confirm the energy extraction capabilities for the offset-slit cylinders using a coil-magnet arrangement, the variation of the non-dimensional extracted average power and the energy transfer ratio over the range of reduced velocities are calculated, which shows the maximum peak power extraction for the case of offset-slit with $\alpha = +30^o$. However, $\alpha = +30^o$ case exhibits a higher peak power than the no-slit case; its peak energy transfer ratio (or energy harvesting efficiency) is similar to the no-slit case, with a relatively wide range of higher efficiency. This makes $\alpha = +30^o$ favorable for energy harvesting applications.

This study leads to developing an efficient VIV-based energy conversion device or improving current flow-induced motion-based converters for low Reynolds number flows. The VIV response and average extracted power increase when the normal slit is set near the front-stagnation point under the lock-in range. Optimal slit location depends on incoming flow circumstances, cylinder structural qualities (mass ratio, damping ratio, etc.), and energy conversion types (e.g., electromagnetic or piezo-electric conversion process). Future studies will address them. Our study illustrates the energy-extraction capabilities of the slit-cylinder from a fundamental perspective; its use in VIV-based harvesters is expected.

**DATA AVAILABILITY**

The data that support the findings of this study are available from the corresponding author upon reasonable request.

**ACKNOWLEDGMENTS**

The authors would like to acknowledge the National Supercomputing Mission (NSM) for providing the computational resources of 'PARAM Sanganak' at IIT Kanpur, which is implemented by C-DAC and supported by the Ministry of Electronics and Information Technology (MeitY) and Department of Science and Technology (DST), Government of India. The authors would also like to acknowledge the IIT-K Computer center (www.iitk.ac.in/cc) for providing the resources to perform the computation work.

**APPENDIX: CALCULATION OF GRID-CONVERGENCE INDEX (GCI)**

The grid-convergence study is performed by taking Grid-3 as the base grid and approximating the error in Grid-4 compared to Grid-3 from the Richardson error estimator, defined by,

$$E_1^{fine} = \frac{\varepsilon_{43}}{1 - r_{43}^0} \qquad (A.1)$$

Error in Grid-2, compared to the solution of Grid-3, is approximated by the coarse-grid Richardson error estimator, which is defined as,

$$E_1^{coarse} = \frac{r^0 \varepsilon_{32}}{1 - r_{32}^0} \qquad (A.2)$$

Where r is the grid refinement ratio between the two consecutive grids defined as,

$$r_{i+1,i} = \frac{h_{i+1}}{h_i}$$

The error ($\varepsilon$) is estimated from the solutions of two consecutive grids by,

$$\varepsilon_{i+1,i} = \frac{f_{i+1} - f_i}{f_i} \qquad (A.3)$$

Grid-Convergence Index (GCI), which accounts for the uncertainty in the Richardson error estimator, is calculated for fine-grid and coarse-grid as given[43-45],

$$GCI_{fine} = F_s \mid E_1^{fine} \mid \qquad (A.4)$$

$$GCI_{coarse} = F_s \mid E_2^{coarse} \mid \qquad (A.5)$$

Further, to calculate the order of accuracy, $L_2$ norm of the errors between is grids is calculated as,

$$L_2 = \sqrt{\left(\sum_{i=1}^{N} \mid \varepsilon_{i+1,i} \mid^2 / N\right)} \qquad (A.6)$$

Where N is the total number of grid points considered for the calculation of $L_2$ norm.